\documentclass[12pt]{iopart}

\usepackage{graphicx}

\begin{document}


\title{On the hyperfine interaction in rare-earth Van Vleck paramagnets
at high magnetic fields }

\author{D Tayurskii\dag\ddag\
\footnote[3]{To whom correspondence should be addressed
(dtayursk@mi.ru)}, H Suzuki\dag}

\address{\dag\ Physics Department, Faculty of Science,
Kanazawa University, Kanazawa, 920-1192, Japan}
\address{\ddag\ Physics Department, Kazan State University, Kazan
, 420008, Russia}

\date{\today}

\begin{abstract}
An influence of high magnetic fields on hyperfine interaction in
the rare-earth ions with non-magnetic ground state (Van Vleck
ions) is theoretically investigated for the case of $Tm^{3+}$ ion
in axial symmetrical crystal electric field (ethylsulphate
crystal). It is shown that magnetic-field induced distortions of
$4f$-electron shell lead to essential changes in hyperfine
magnetic field at the nucleus. The proposed theoretical model is
in agreement with recent experimental data.
\end{abstract}

\pacs{76.60.-k, 76.30.Kg, 75.10.Dg}


\section{Introduction}
Van Vleck (VV) or polarization magnetism most often occurs in
crystals containing non-Kramers rare-earth (RE) ions, i.e. ions
with an even numbers of electrons in the unfilled $4f$-shell,
where the crystalline electric field lifts the degeneracy of the
ground multiplet $^{2S+1}L_J$, leading to typical splittings of
the Stark structure of the order of 10-100 cm$^{-1}$. The
electronic ground state in this case is a singlet or non-magnetic
doublet so all magnetic properties of Van Vleck paramagnets are
necessitated by the Zeeman effect which at moderate magnetic
fields (the Zeeman energy is much more less then the Stark
splittings) can be calculated in the second order of perturbation
theory \cite{VV,AB,Teplov}. The rather strong hyperfine
interaction induces the magnetic field at the nucleus of VV ion in
many times greater than the applied external magnetic field and
leads to enormous values (up to several hundred) of paramagnetic
shift of the NMR lines. This so-called "enhanced" NMR is one the
most important methods for studying the magnetic properties of VV
paramagnets \cite{AB,Teplov}.

At high magnetic fields the Zeeman energy of RE ion becomes
comparable with the Stark splitting energies and a number of new
physical effects appears \cite{Rev}. Among them the magnetic field
induced structural phase transitions in dielectric VV paramagnets
$TmPO_4$ \cite{TmPO4} and $LiTmF_4$ \cite{LiTmF4} and the coupled
$4f$-electron-phonon excitations in the thulium ethylsulphate
crystal $Tm(C_2H_5SO_4)_2 \cdot 9H_2O$ ($TmES$)\cite{elphon} can
be mentioned. From the theoretical point of view the applicability
of perturbation theory is violated at high magnetic fields and a
new theory has to be built \cite{Rev,EPR}. For example, it was
shown in \cite{Rev,elnucl} that at high magnetic fields the
coupled $4f$-electron-nuclear states in insulating VV paramagnets
appear. The resonant absorption due to the transitions between
electronic-nuclear sublevels of the ground singlet in $TmES$
crystal has been observed in \cite{NMR} at magnetic fields up to
$3 T$. It should be pointed out that the frequencies of those
transitions lie almost in the X band of the EPR frequencies while
the transition probabilities are determined by the matrix elements
of the nuclear spin operator. From this point of view it is
reasonable to speak about "ultrahigh- frequency" NMR at high
magnetic field, in contrast to the "enhanced" NMR at moderate
magnetic fields. The observed field dependence of transition
frequencies doesn't coincide with the predicted one in
\cite{Rev,elnucl} where for clarity in setting forth the essential
changes arising in properties of the nuclear-spin system of an
insulating VV paramagnet under influence of a high magnetic field
the possible changes in hyperfine interaction parameters were
neglected. But a sufficiently high magnetic field will cause
distortion of the $4f$-electron shell and a redistribution of the
electron density, inevitably altering the hyperfine field at the
nucleus.

In this Paper, we investigate theoretically the influence of a
rather high external magnetic fields on hyperfine interaction in
dielectric VV paramagnets. As a model system the well-studied at
moderate magnetic fields $TmES$ crystal is considered.

\section{Hyperfine interaction in VV paramagnets}
The Hamiltonian of an isolated VV ion (the distance between two
nearest-neighbor thulium ions in $TmES$ is $\sim7$ $\AA$ and
single-ion approximation works a rather well) can be written as
\begin{equation}
\mathcal{H}=\mathcal{H}_{cr}+\mathcal{H}_{eZ}+\mathcal{H}_{nZ}+
\mathcal{H}_{hf},\label{eq:1}
\end{equation}
where the crystal electric field Hamiltonian in generally accepted
notations \cite{ABbook} reads as
\begin{equation}
\mathcal{H}_{cr}=\alpha B_{20}O^{0}_{2}+\beta
B_{40}O^{0}_{4}+\gamma
(B_{60}O^{0}_{6}+B_{66}O^{6}_{6}).\label{eq:2}
\end{equation}
The Hamiltonian of the Zeeman interaction of $4f$-electron shell
and that of the nuclear Zeeman interaction are in usual forms:
\begin{equation}
\mathcal{H}_{eZ}=g_{J}\mu_{B}\mathbf{H}\mathbf{J},
\mathcal{H}_{nZ}=-\gamma_{I}\hbar\mathbf{H}\mathbf{I}.\label{eq:3}
\end{equation}
The explicit form of the Hamiltonian of hyperfine interaction
$\mathcal{H}_{hf}$ can be estimated in the case of RE ions in the
following way (see, for example, \cite{ABbook}; note, that we
shall not consider hereinafter quadrupole effects because of
$^{169}Tm$ nuclear spin is one-half). For a free RE the magnetic
interaction with the nucleus of the electrons in the partially
filled $4f$-shell with orbital moments $\mathbf{l}_i$ and spins
$\mathbf{s}_i$ is given by
\begin{equation}
\mathcal{H}_{hf}=2\mu_{B}\gamma_{I}\hbar\sum_{i\in 4f}\{r_{i}^{-3}
[\mathbf{l}_{i}-\mathbf{s}_{i}+3\mathbf{r}_{i}(\mathbf{r}_{i}\mathbf{s}_{i})
/r_{i}^{2}]\}\cdot \mathbf{I}= 2\mu_{B}\gamma_{I}\hbar\langle
r_{i}^{-3}\rangle (\mathbf{N}\cdot\mathbf{I})\label{eq:4}
\end{equation}
and can be represented within the the ground multiplet manifold
$^{2S+1}L_{J}$ where the total angular momentum $\mathbf{J}$ is
the good quantum number as
\begin{equation}
\mathcal{H}_{hf}=2\mu_{B}\gamma_{I}\hbar\langle r_{i}^{-3}\rangle
\langle J \|N\|J\rangle
(\mathbf{J}\cdot\mathbf{I})=A_{hf}(\mathbf{J}\mathbf{I}).\label{eq:5}
\end{equation}
The reduced matrix elements $\langle J \|N\|J\rangle$ are
tabulated for different $4f^n$ electronic configurations
\cite{ABbook}.

For an RE ion in a crystal the crystal electric field reduces the
rotational symmetry of free atom and removes partially or
completely the degeneracies of energy levels. As the result there
asymmetry reflected the local environment symmetry of an RE ion
appears in hyperfine interaction parameter $A_{hf}$. So the
Eq.\ref{eq:5} has to be replaced by
\begin{equation}
\mathcal{H}_{hf}=(\mathbf{J}\tilde{A}\mathbf{I}).\label{eq:6}
\end{equation}
For example, in the ethylsulphates of RE ions with magnetic ground
states (electronics levels exhibit Kramer's degeneracy) the
principal values of the hyperfine interaction tensor $\tilde{A}$
obtained by measurements of hyperfine structure of the
paramagnetic resonance lines can differ in $10$ times and more.
But for VV ions, i.e. non-Kramer's ions with a non-magnetic ground
state like $Tm^{3+}$ in $TmES$ (the total angular momentum is
quenched by the crystal electric field), the paramagnetic
resonance is not observable and the principal values of
$\tilde{A}$ are determined by indirect way. For our estimations we
have used the measured by means of the enhanced NMR at moderate
magnetic fields so-called paramagnetic shift \cite{PS} which is
anisotropic and the values of which depend on the strength of
hyperfine interaction as well as on the degree of mixing of the
Stark wave functions by the applied magnetic field. Taking into
account the explicit form of wave functions for the Hamiltonian
$\mathcal{H}_{cr}+\mathcal{H}_{eZ}$ we get after calculations the
following principal values of the hyperfine interaction tensor for
$Tm^{3+}$ in the axial crystal field in $TmES$:
$A_{\parallel}\approx -241$\,MHz and $A_{\perp}\approx -388$\,MHz
(the estimated value of $A_{hf}$ for free tripositive ion
$Tm^{3+}$ from measurements of $Tm$ and $Tm^{2+}$ is -393.5\,MHz
\cite{ABbook}).

A rather high magnetic field the Zeeman energy in which is
comparable with the Stark splittings of the ground multiplet leads
to the further lowering of symmetry and consequently to the
redistribution of $4f$-electron density in VV ions. The inclusion
of a such magnetic field can be considered formally as the
appearance of low-symmetry magnetic field  dependent term in the
crystal field Hamiltonian. Consequently we can assume that
magnetic hyperfine interaction is still described by Eq.\ref{eq:6}
where the components of hyperfine interaction tensor $\tilde{A}$
depend on the strength of applied magnetic field.

Up to our knowledge \emph{ab initio} calculations of  crystal
field effects and magnetic properties have not been provided for
the VV paramagnets yet. So in order to estimate theoretical values
of hyperfine interaction parameters in insulating VV paramagnets
at high magnetic field and to compare the obtained results with
the observed in \cite{NMR} magnetic field dependence of the
resonant frequencies due to transitions between electron-nuclear
states in $TmES$ the following phenomenological model can be
proposed. Assuming that the external magnetic field is not too
high to destroy partially \emph{LS} coupling and taking into
account that hyperfine magnetic field at nucleus is determined by
the magnetic moment of $4f$-electron shell one can write for the
the principal values of hyperfine interaction tensor at high
magnetic field the following relation:
\begin{equation}
\frac{A_{\parallel,\perp}(H)}{M_{\parallel,\perp}(H)}=
\frac{A_{\parallel,\perp}}{M_{\parallel,\perp}},\label{eq:8}
\end{equation}
where $M_{\parallel,\perp}(H)$ and $M_{\parallel,\perp}$
represents the components of the electronic magnetic moment of
$Tm^{3+}$ ion at high magnetic field and at moderate magnetic
field respectively, and can be calculated by use the wave
functions of the Hamiltonian $\mathcal{H}_{cr}+\mathcal{H}_{eZ}$.
The principal values $A_{\parallel,\perp}$ have been determined
above. Note here that because of dependence of $4f$-electron
magnetic moment in VV paramagnets on the populations of the
nearest excited levels (at helium temperatures) the temperature
dependence of hyperfine interaction parameters is expected. The
provided calculations with use above model allowed to describe the
observed experimental data \cite{NMR} in a good way.
\section{Summary}

In summary, we have showed that at rather high magnetic field the
Zeeman energy of an Van Vleck ion in which is comparable with
Stark structure splitting energies the components of magnetic
hyperfine interaction tensor begin to be dependent on the applied
magnetic field strength. The phenomenological model for taking
into account the influence of the induced external magnetic field
distortions of $4f$-electron shell on hyperfine magnetic field at
nucleus in insulating VV paramagnets has been proposed. The
results of these studies can be applied also to intermetallic VV
paramagnets which are very interesting from the point of view of
ultra-low temperature physics.


The authors thank M.S.Tagirov (Kazan State University) for
discussions.



\section*{References}

\begin{thebibliography}{99}

\bibitem{VV}
Van Vleck J H 1929 {\it Phys. Rev.} {\bf 34} 1494

\bibitem{AB}
Abragam A and Bleaney B 1983 {\it Proc. R. Soc. London, Ser. A}
{\bf 387} 221

\bibitem{Teplov}
Aminov L K and Teplov M A 1985 {\it Sov. Phys. Usp.} {28} 762\\
\hspace*{-3.4mm}Aminov L K and Teplov M A 1990 {\it Sov. Sci. Rev.
A, Phys. Rev} {\bf 14} 1 and references therein

\bibitem{Rev}
Tagirov M S and Tayurskii D A 2002 {\it Low Temp. Phys.} {\bf 28}
147

\bibitem{TmPO4}
Vekhter B G {\it et al.} 1991 {\it JETP Lett.} {\bf 54} 578

\bibitem{LiTmF4}
Klochkov A V {\it et al.} 1997 {\it JETP Lett.} {\bf 66} 266

\bibitem{elphon}
Tayurskii D A and Tagirov M S 1998 {\it JETP Lett.} {\bf 67} 1040

\bibitem{EPR}
Tagirov M S and Tayurskii D A 1995 {\it JETP Lett.} {\bf 61} 672

\bibitem{elnucl}
Tayurskii D A {\it et al.} 2000 {\it Physica B} {\bf 284-288} 1686

\bibitem{NMR}
Abubakirov D I {\it et al.} 2002 {\it JETP Lett.} {\bf 76}
accepted

\bibitem{ABbook}
Abragam A and Bleaney B {\it Electron Paramagnetic Resonance of
Transition Ions} 1970, Clarendon Press, Oxford

\bibitem{WF}
Watson R E and Freeman A J 1964 {\it Phys. Rev.} {\bf 133} A1571

\bibitem{PS}
Abdulsabirov R Yu {\it et al.} 1979 {\it Sov. Phys. JETP} {\bf 49}
517






\end{thebibliography}
\end{document}